\documentclass[12pt]{article}
\usepackage{amsmath}
\usepackage{epsf}
\usepackage{epsfig}
\usepackage{here}
\usepackage{amssymb}
\usepackage{citesort}
\usepackage{graphicx}
\usepackage{latexsym}
\newcommand{\be}{\begin{equation}}
\newcommand{\ee}{\end{equation}}
\newcommand{\bear}{\begin{eqnarray}}
\newcommand{\ear}{\end{eqnarray}}

\newsavebox{\LSIM}
\sbox{\LSIM}{\raisebox{-1ex}{$\ \stackrel{\textstyle<}{\sim}\ $}}
\newcommand{\lsim}{\usebox{\LSIM}}
\newsavebox{\GSIM}
\sbox{\GSIM}{\raisebox{-1ex}{$\ \stackrel{\textstyle>}{\sim}\ $}}
\newcommand{\gsim}{\usebox{\GSIM}}
\newcommand{\vev}{\langle \phi\rangle}
\newcommand{\MP}{M_{\rm Pl}}

\begin{document}
\begin{titlepage}
\begin{flushright}
BA-02-23\\
DESY 02-098\\
hep-ph/0207232
\end{flushright}
$\mbox{ }$
\vspace{.1cm}
\begin{center}
\vspace{.5cm}
{\bf\Large Cosmological Constant, Gauge Hierarchy}\\[.3cm]
{\bf\Large And Warped Geometry}\\
\vspace{1cm}
Stephan J. Huber$^{a,}$\footnote{stephan.huber@desy.de}
%Chin-Aik Lee$^{a,}$\footnote{jlca@udel.edu} 
and 
Qaisar Shafi$^{b,}$\footnote{shafi@bartol.udel.edu} \\ 
 
\vspace{1cm} {\em  
$^a$Deutsches Elektronen-Synchrotron DESY, Hamburg, Germany}\\[.2cm] 
{\em $^b$Bartol Research Institute, University of Delaware, Newark, USA} 

\end{center}
\bigskip\noindent
\vspace{1.cm}
\begin{abstract}
It is suggested that the mechanism responsible for the resolution of the  
gauge hierarchy problem within the warped geometry framework can be
generalized to provide a new explanation of the extremely tiny 
vacuum energy density $\rho_V$ suggested by recent observations. We illustrate
the mechanism with some 5D examples in which the true vacuum energy is
assumed to vanish, and $\rho_V$ is associated with a false vacuum energy such
that $\rho_V^{1/4}\sim {\rm TeV} ^2/ M_{\rm Pl} \sim 10^{-3}$ eV, 
where $M_{\rm Pl}$ denotes the reduced Planck mass. We also consider
a quintessence-like solution to the dark energy problem.   
\end{abstract}
\end{titlepage}
\section{Introduction}
Evidence for a tiny but non-zero cosmological vacuum energy density has
steadily mounted in recent years \cite{CC_OBS}. With information from a variety  
of observations put together, it appears that we are living in a flat
universe, such that $\Omega_{\Lambda} + \Omega_m =1$, where 
$\Omega_{\Lambda} = 0.7$ and $\Omega_m = 0.3$ denote the 
density parameters associated with the
vacuum energy and matter respectively. These observations pose at least
two formidable theoretical challenges, namely, what physics determines
the magnitude of the vacuum energy density to be of order
\begin{equation} \label{lambda}
\rho_V\sim 10^{-120}\MP^4\sim(10^{-3}{\rm ~eV})^4,
\end{equation}
and why are $\rho_V$ and $\rho_m$ in
such close proximity in magnitude, considering that their evolution, at
least for constant cosmological constant, can be so different. Here 
$\MP= 2.4\cdot10^{18}$ GeV denotes the reduced Planck mass.
  
A certain amount of theoretical prejudice suggests
that the true vacuum energy density of the universe
could be exactly zero.
Needless to say, how the zero value occurs remains a complete
mystery, despite many attempts \cite{EV}. Nonetheless,
following this reasoning, it has been speculated that the present universe
happens to lie in a false vacuum, separated from the true vacuum by the
observed energy density $\rho_V$ \cite{KKS}. The modest goal of this approach
is to try to identify the physics underlying the origin of $\rho_V$, as well as
ensure that the metastable vacuum is sufficiently long lived.
   
In this letter we propose a new mechanism for realizing $\rho_V$ of the
desired magnitude which is based on warped geometry. While the original
motivation of warped geometry was to resolve the notorious gauge
hierarchy problem \cite{RS,G}, it has subsequently been exploited to provide at
least a qualitatively new understanding of fermion mass hierarchies and their    
mixings, especially within the context of neutrino oscillations \cite{GN,GP,HS2,HS3}.
We will show here that the warped geometry can be further exploited
to provide a new mechanism for generating $\rho_V$ of the desired
magnitude. Earlier attempt in this direction have been made in ref.~\cite{TWCF}. 
Recall that the TeV scale in warped geometry arises from the 
presence of the warp factor $\Omega^{-1}=\exp(-\pi k R) = {\rm TeV}/ M_P$, 
where $1/k$ denotes the AdS curvature radius and $R$ is the radius of the
orbifold $S^1/ Z_2$. We will see that $\rho_V$ can be associated with
the vacuum energy density of a suitable scalar field, such that $$\rho_V^{1/4}\sim
\Omega^{-2}\MP = {\rm TeV}^2/ \MP \sim 10^{-3} {\rm eV}.$$ 
We also discuss how the quintessence scenario \cite{quin} can be realized within
this approach.

\section{Scalar Fields in a Warped Background}
We take the fifth dimension to be an $S_1/Z_2$ orbifold with 
a negative bulk cosmological constant, bordered by two 3-branes 
with opposite tensions and separated by distance $R$.  Einstein's
equations are satisfied by the non-factorizable metric  \cite{RS}
\begin{equation} \label{met}
ds^2=e^{-2\sigma(y)}\eta_{\mu\nu}dx^{\mu}dx^{\nu}+dy^2, ~~~~\sigma(y)=k|y|
\end{equation}
which describes a slice of AdS$_5$. The 4-dimensional metric is 
$\eta_{\mu\nu}={\rm diag}(-1,1,1,1)$, $k$ is the AdS curvature 
related to the bulk cosmological constant and brane tensions, 
and $y$ denotes the fifth coordinate. The AdS curvature and the 
5d Planck mass $M_5$ are both assumed to be of order $M_{\rm Pl}$.
The AdS warp factor $\Omega=e^{\pi k y}$
generates an exponential hierarchy of energy scales.
If the brane separation is $kR\simeq 11$, the scale at
the negative tension brane, located at $y=\pi R$, is of TeV-size, 
while the scale at the brane at $y=0$ is of order $M_{\rm Pl}$. 
At the TeV-brane gravity is weak because the zero mode
corresponding to the 4D graviton is localized at the
positive tension brane (Planck-brane). 

Let us consider the equation of motion of a real scalar field
in the background (\ref{met}) \cite{GW,GP}
\begin{equation}
\frac{1}{\sqrt{-g}}\partial_M(\sqrt{-g}g^{MN}\partial_N\Phi)-M^2\Phi=0
\end{equation}
which follows from the action
\begin{equation} \label{action}
S_5=-\int d^4x \int dy~\sqrt{-g}\left(\frac{1}{2}(\partial_M\Phi)(\partial^M\Phi)+\frac{1}{2}
M^2\Phi^2\right).
\end{equation}
Here $g_{MN}$ denotes the 5D metric and $g=e^{-4\sigma}$ its determinant. 
In general, the 5D scalar mass consists of bulk and brane contributions
\begin{equation}
M^2(y)=b^2k^2+a^2k\delta(y-\pi R)+\tilde{a}^2k\delta(y),
\end{equation}
where $b$, $a$ and $\tilde a$ are dimensionless parameters.
In the case of supersymmetry the relations
\begin{eqnarray} \label{susy}
\tilde{a}^2&=&-a^2 \nonumber \\
b^2&=&\frac{1}{4}a^2(a^2+8)
\end{eqnarray}
have to be satisfied \cite{GP}, and  $\Phi$ becomes
a complex scalar field which, together
with a second complex scalar and a Dirac fermion, forms
a hypermultiplet. The connection to the 5D mass parameter
of the Dirac fermion $M_{\Psi}=c\sigma'$ is given by $c=3/2+a^2/2$.
A special case of eqs.~(\ref{susy}) is the massless scalar field
where $a^2=\tilde{a}^2=b=0$ (and $c=3/2$).

The effective 4D theory arises from the Kaluza-Klein (KK)
decomposition
\begin{equation}
\Phi(x^{\mu},y)=\frac{1}{\sqrt{2\pi R}}\sum_{n=0}^{\infty}\Phi^{(n)}(x^{\mu})f_n(y),
\end{equation}
where the wave functions $f_n(y)$ satisfy the differential
equation \cite{GW,GP}
\begin{equation} \label{eom}
(-\partial_5^2+4\sigma'\partial_5+M^2)f_n=
        e^{2\sigma}m_n^2f_n,
\end{equation} 
and $m_n$ are the masses of the KK excitations.
The solution to eq.~(\ref{eom}) is given by 
\begin{equation} \label{f1}
f_n(y)=\frac{e^{2\sigma}}{N_n}\left[J_{\alpha}(\frac{m_n}{k}e^{\sigma})+
                 \beta_{\alpha}(m_n)Y_{\alpha}(\frac{m_n}{k}e^{\sigma})\right],
\end{equation}
with $\alpha=\sqrt{4+b^2}$. Assuming $\Phi$ is even under the $Z_2$
orbifold transformation, i.e.~$f_n(-y)=f_n(y)$, the coefficient $\beta$ 
and the KK spectrum follow from the matching conditions at the 
branes, 
\begin{eqnarray} \label{quant1}
 \beta_{\alpha}(x_n,\tilde a^2)&=&-\frac{(-\tilde a^2+4-2\alpha)
                            J_{\alpha}(x_n)
                            +2x_nJ_{\alpha-1}(x_n)}
                           {(-\tilde a^2+4-2\alpha)Y_{\alpha}
                            (x_n)
                            +2x_nY_{\alpha-1}(x_n)}, 
\\[2mm]  \label{quant2}
 \beta_{\alpha}(x_n,\tilde a^2)&=&\beta_{\alpha}(\Omega x_n,-a^2),
\end{eqnarray}
where  $x_n=m_n\Omega/k$.
Note that for non-vanishing boundary mass terms the derivative
of $f_n$ becomes discontinuous on the boundaries. The normalization
constants $N_n$ in (\ref{f1}) are defined such that 
\begin{equation} \label{norm}
\frac{1}{2\pi R}\int_{-\pi R}^{\pi R}dye^{-2\sigma(y)}f_n^2(y)=1. 
\end{equation}
The mass
splitting between the KK modes is $\Delta m_n\sim\pi k \Omega^{-1}\sim$ TeV.

In the massless case the zero mode wave function is constant, 
i.e.~$f_0(y)=1/N_0\approx \sqrt{2\pi kR}$. The KK states have TeV
scale masses and are localized at the TeV-brane. The mass of
the first excited state is $m_1\approx 3.83k\Omega^{-1}$.
The effects of the bulk and brane mass terms on the mass and
wave function of the lowest mode are quite different. If one switches on
a small bulk mass, one finds $m_0\approx bk/\sqrt{2}$ for 
$b\lsim \Omega^{-1}$~($bk\lsim$ TeV). Thus, there is no
warp factor suppression. The zero mode wave function is still
approximately constant, with a small enhancement at the TeV-brane.
Increasing the bulk mass to $b\lsim1$ ($bk\lsim M_{\rm Pl}$)
the mass of the lightest mode remains frozen at $x_0=3.83$,
which is practically the former first excited state. For   
$b\gsim1$ ($bk\gsim M_{\rm Pl}$) one finds that $m_0$ also increases.
The mass term at the  Planck-brane induces a similar
behavior of the zero mode, except there is no  increase of $m_0$ for 
$\tilde{a}\gsim1$ ($\tilde{a}k\gsim M_{\rm Pl}$).  

\begin{figure}[t] 
\begin{picture}(100,170)
\put(25,5){\epsfxsize6.5cm \epsffile{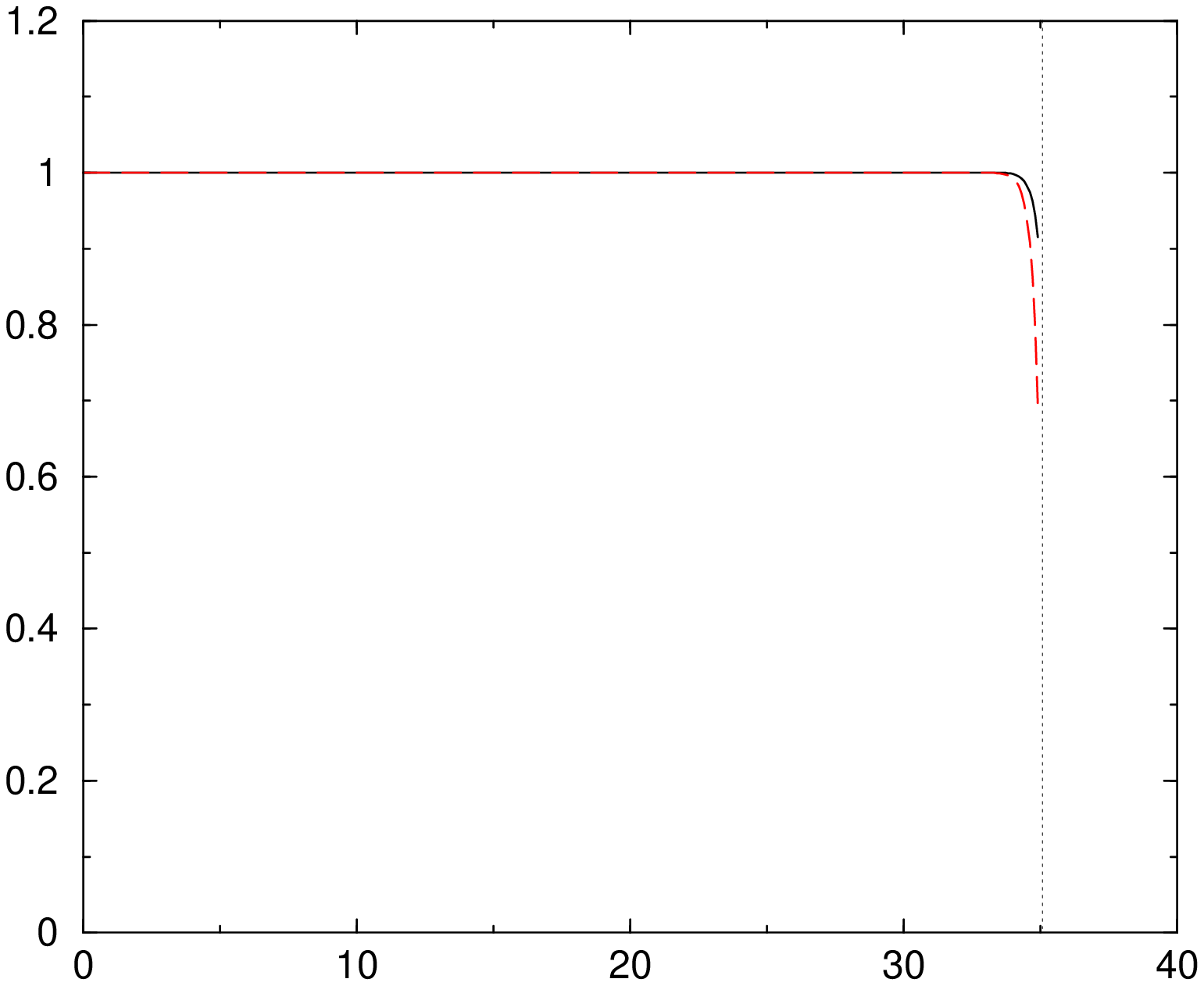}}
\put(250,5){\epsfxsize6.2cm \epsffile{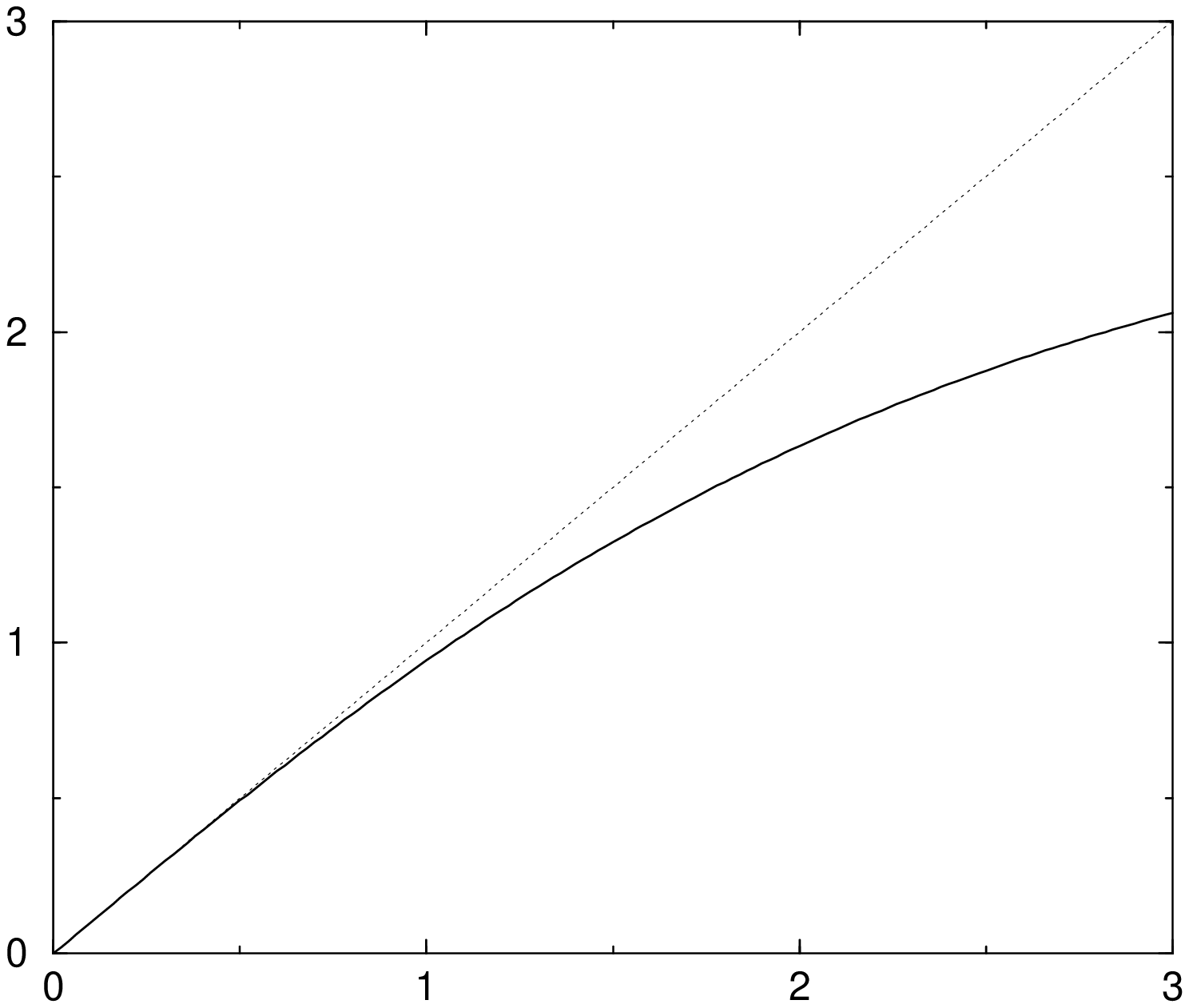}}
\put(-15,130){{$\frac{f_0}{\sqrt{2\pi kR}}\uparrow$}} 
\put(215,130){{$\frac{m_0}{k}\Omega^2\uparrow$}} 
\put(155,-5){$ky\rightarrow$}
\put(385,-5){$a\rightarrow$}
\end{picture} 
\caption{(a): Wave function of the zero mode for $a=1$ (solid line)
and $a=2$ (dashed line). (b): Dependence of the zero mode mass
on the TeV-brane mass term (solid line) compared to the linear
approximation of eq.~(\ref{lin}) (dotted line).
}
\label{f_1}
\end{figure}

In the case of a TeV-brane mass the result is very different.
Because of the factor $\sqrt{-g}$ in the action (\ref{action}),
the mass of the zero mode is warped by {\em two} powers of 
$\Omega$. With this  double warping we have
\begin{equation} \label{lin}
m_0\approx ak\Omega^{-2},~~ a\lsim1. 
\end{equation}
The wave function is nearly constant, with a dip at the 
TeV-brane, as shown in fig.~\ref{f_1}a for $kR=11.13$ 
($M_{\rm Pl}e^{-\pi kR}=1.6$ TeV).
For $a\gsim1$ the dip in the wave function slows down the
increase of $m_0$. For large values of $a$ any further increase
is completely compensated by this, so that the zero mode
mass saturates at $m_0=2\sqrt{2}k\Omega^{-2}$.
The dependence of $m_0$ on the
TeV-brane mass is displayed in fig.~\ref{f_1}b.
We note that gauge bosons show a similar behavior
as we discussed in ref.~\cite{HS}. 
Numerically, for a TeV-brane mass of order $M_{\rm Pl}$, $m_0$ is 
comparable to the energy scale associated with  dark energy (\ref{lambda}).
Of course, at this stage $m_0$ is not yet related to some vacuum
energy, but merely represents a tiny particle mass. In the next section we will
show how such a link can be established. 

At the end of this section, let us briefly summarize the supersymmetric
case which was extensively discussed in refs.~\cite{GP,GP2}. 
Supersymmetry requires the bulk and brane mass terms to be 
simultaneously present. Because of the special relations 
eq.~(\ref{susy}), the zero mode remains strictly massless, despite
the non-zero bulk and brane masses. The wave function has 
an exponential form \cite{GP},
\begin{equation} \label{wave_susy}
f_0(y)=\frac{1}{N_0}e^{-\frac{a^2}{2}\sigma}.
\end{equation}
For a positive (negative) mass squared term at the TeV-brane the scalar field is
localized towards the Planck-brane (TeV-brane). The supersymmetric
scalar fields thus behave similar to fermions in the warped background, which
also possess massless exponential zero modes in the presence of bulk mass
terms \cite{GN,GP}.

\section{Exponentially Small Contributions to the Scalar Potential}
In the last section we learned that a Planck-size mass term at
the TeV-brane is translated into a sub-eV mass of the scalar particle.
We now generalize this result to include a scalar potential on the
TeV-brane. This will allow as to generate tiny energy splittings
in the minima of the 4D effective potential, which we hope to identify
with the observed dark energy.

We start with the simplest case of a massless scalar field with $a=b=\tilde{a}=0$
which, for instance, could be a consequence of supersymmetry.
If we now switch on some general potential on the TeV-brane, 
related for example, to supersymmetry breaking,
the scalar zero mode remains, as discussed in the last section.
Consider the following contribution to the 5D action
\begin{equation} \label{brane}
\Delta S_5=-\int d^4x \int dy~\sqrt{-g}\left(\frac{1}{2}
a^2k\Phi^2+\frac{1}{3}E_5k^{-1/2}\Phi^3+\frac{1}{4}\lambda_5k^{-2}\Phi^4\right)
\delta(y-\pi R).
\end{equation}
We approximate $f_0$ by a constant, and after
integrating over the extra dimension, we arrive at
the 4D effective potential for the zero mode $\phi$
\begin{equation} \label{potential}
V_4=\frac{1}{2}m^2\phi^2+\frac{1}{3}E\phi^3+\frac{1}{4}\lambda\phi^4,
\end{equation}
where
\begin{eqnarray} \label{couplings}
%L&=&L_5k^3\Omega^{-4}\nonumber\\
m^2&=& a^2k^2\Omega^{-4}\nonumber\\
E&=&E_5k\Omega^{-4}\nonumber\\
\lambda&=&\lambda_5\Omega^{-4}.
\end{eqnarray}
Thus all operators on the TeV-brane are
red-shifted by four powers of the warp factor.
The warped geometry not only generates
tiny masses but also exponentially small
coupling constants. This result may seem surprising
since the zero mode $f_0$ is constant in the
extra dimension and thus has a large overlap with the
TeV-brane. However, as we observe from 
eq.~(\ref{norm}), in the integration over the extra dimension
the region close to the TeV-brane is exponentially suppressed.
The rescaled zero mode $\tilde{f}_0=e^{-\sigma}f_0$ is the
more appropriate object to look at. For a constant $f_0$,
the rescaled zero mode is exponentially localized towards
the Planck-brane, which explains the tiny quantities in
eq.~(\ref{couplings}). For a very large mass squared term on the
TeV-brane, i.e.~$a\gg1$, there is an additional suppression
of the brane operators (\ref{brane}) induced by the dip in
the zero mode. We note that scalar couplings in the bulk or on the
Planck-brane are not scaled down by the warped geometry.

We now assume that $m^2<0$ in order to generate
a vev for the scalar field. Without additional
interactions from the bulk, the tiny self
coupling $\lambda$ induced by the brane can stabilize the field
only at a Planck-size vev $\vev\sim k$.
We therefore have to introduce some additional interaction.
A quartic coupling in the bulk does not feel
the warping, and therefore generates $\lambda\sim1$
in the effective 4D action. A quartic coupling on the 
Planck-brane would have the same effect.
The scalar field then acquires
a tiny vev $\vev\sim k\Omega^{-2}$. If there is now
more than one minimum with a similar vev, the energy 
splitting between the two minima will be on the order of 
$\Delta V\sim m^2 \vev^2 \sim k^4\Omega^{-8}$. 
This is exactly what is required for the
dark energy. Two minima can easily arise, for example,  
in the case of two interacting scalar fields with a 
polynomial potential. 
We will show later that the false vacuum has
a lifetime much longer than the age of the universe. 
In the next section we will also discuss how this setup can
be stabilized against radiative corrections in a supersymmetric
framework.

A different possibility for the scalar field to be
stabilized is by some non-renor\-malizable interaction
in the bulk, for instance, by a $(1/Q^2)\phi^6$ operator.
As discussed in refs.~\cite{GP,HS2}, non-renormalizable
interaction may be suppressed by much smaller energy scales
than $M_{\rm Pl}$ because of the warped geometry. 
However, as explained above, in some sense the flat zero 
mode is localized towards the Planck-brane. Therefore 
non-renormalizable operators are not enhanced, and we
expect $Q\sim M_{\rm Pl}=k$ in the 4D action. The scalar would
then be stabilized at $\vev=\sqrt{mQ}\sim k\Omega^{-1}$, which
is the KK scale. Assuming that the $Z_2$ symmetry in the
scalar potential is broken by the cubic term on the brane,
the induced energy splitting between the vacua at $\pm\sqrt{mQ}$ is 
$\Delta V\sim E \vev^3 \sim k^4\Omega^{-7}$. This is a
factor of $\Omega$ larger than what we found in the 
previous example. To make it compatible with the observational
value (\ref{lambda}) we either would have to increase the warp
factor and end up with a KK scale of about 10 GeV(!), or to reduce 
the suppression scale to about $Q=k\Omega^{-1/3}$.
Both possibilities do not look realistic.
In the next section we discuss the supersymmetric 
case (with a non-vanishing bulk mass) which gives us 
additional possibilities.

The false vacuum energy not only leads to an exponential 
expansion of the universe, but will also modify the geometry 
of the extra dimension. To estimate the size of the back reaction
we compare the false vacuum energy density to the
bulk cosmological constant and brane tensions of the unperturbed
Randall-Sundrum solution \cite{RS}. We consider a somewhat
simplified setup where the scalar vev is induced by a 
negative squared mass term $(1/2)a^2k\Phi^2$ at the TeV-brane,
as in eq.~(\ref{brane}). The potential is stabilized by a
quartic self interaction in the bulk, $(1/4)\lambda_{\rm b}\Phi^4$.
Because the scalar potential changes along the extra dimension, 
the vacuum configuration $\langle \Phi(y) \rangle$ does as well and
has to be computed numerically. We find the profile to be 
$\langle \Phi(y) \rangle\sim ak\Omega^{-2}/\sqrt{\lambda_{\rm b}}\sim k^{3/2}\Omega^{-2}$ 
and almost flat in the extra dimension. In the second step 
we assumed $a\sim 1$ and $\lambda_{\rm b}\sim k^{-1}$.
Close to the TeV-brane, where the instability occurs,  
the profile $\langle \Phi(y) \rangle$ increases by about 10 percent
relative to its almost constant value in the bulk.
On the TeV-brane the scalar potential is
$V=(1/2)a^2k\Phi^2 \sim k^4\Omega^{-4}$. In the bulk we
obtain $V=(1/4)\lambda_{\rm b}\Phi^4\sim k^5\Omega^{-8}$.
This means that Einstein's equations are still
completely dominated by Planck-size bulk cosmological constant
and brane tensions of the Randall-Sundrum solution \cite{RS}.
Thus we can safely ignore the back reaction, at most of order $\Omega^{-4}$,
on the wave functions (\ref{f1}) and 
(\ref{wave_susy}).

\section{The Supersymmetric Case}
A crucial point is that a perturbation of the
mass relations (\ref{susy}) by some supersymmetry
breaking mass term on the TeV-brane with $\Delta a^2\lsim1$
leaves the zero mode (\ref{wave_susy}) 
practically unchanged. For $\Delta a^2\gsim1$ we observe
an additional reduction of $f_0$ close to the TeV-brane, 
as in the case of a vanishing bulk mass. To obtain
the 4D couplings from the brane action (\ref{brane}) we
approximate the zero mode by its supersymmetric
shape  (\ref{wave_susy}). We find
\begin{eqnarray} \label{couplings_susy}
m^2&\approx& \Delta a^2k^2\Omega^{-4-a^2}\nonumber\\
E&\approx&E_5k\Omega^{-4-\frac{3}{2}a^2}\nonumber\\
\lambda&\approx&\lambda_5\Omega^{-4-2a^2},
\end{eqnarray}
where $a^2$ is now the TeV-brane mass term required
by supersymmetry and determines the shape of the 
zero mode. We have omitted a factor $N_0\sqrt{2\pi kR}$
to some power which is of order unity. Localizing the 
scalar zero mode towards the Planck-brane by taking $a^2>0$,
we can suppress the induced 4D couplings even stronger
than in eq.~(\ref{couplings}). This opens up new possibilities
to generate a small vacuum energy density.

The ($N=2$) 5D supersymmetry prohibits a self interaction
of the scalar field in the bulk. Self interactions can be provided,
however, by superpotentials on the branes, where the 
boundary conditions break the supersymmetry to $N=1$.
Planck-brane operators are not warped down and
induce order unity couplings in the 4D potential, in contrast
to their TeV-brane counterparts (\ref{couplings_susy}).

Localizing the scalar field close to the Planck-brane, we
can implement a quint\-essence-like solution to the dark energy
problem. The cosmological evolution is assumed to leave the scalar field 
displaced from its minimum towards which it is slowly evolving.
For polynomial potentials,  the necessary slow rolling conditions 
are satisfied only for very large vevs $\vev\sim M_{\rm Pl}$.
A small vacuum energy can therefore be explained only by
extraordinarily small couplings. 
To generate $\rho_V$ from a term  $m^2\vev^2\sim m^2M_{\rm Pl}^2$ 
requires $m^2\sim \Omega^{-8} M_{\rm Pl}^2$. From  
eq.~(\ref{couplings_susy}) we see that this requires
$a^2\approx4$.
We can also implement this scenario by relying on the quartic 
term of the brane potential, $\rho_v\sim\lambda \vev^4$. 
For $a^2\sim2$ we obtain the required small $\lambda \approx \Omega^{-8}$.
The other terms in eq.~(\ref{brane}) as well as possible bulk
self interactions have to vanish to accommodate this solution. 
It remains to be seen whether a realistic scenario of this type can 
be constructed in which the required small masses 
remain stable
under quantum corrections \cite{DV}.

A true cosmological constant, instead of quintessence, can be 
generated from a supersymmetry breaking cubic coupling which 
shifts previously degenerate minima.
We assume that a scalar self interaction on the Planck-brane 
leads to $\lambda\sim1$ in the 4D action. If the scalar potential
is now destabilized by a supersymmetry breaking brane mass 
(\ref{couplings_susy}), the induced vev is 
$\vev\sim m\sim k\Omega^{-2-a^2/2}$. The degeneracy of the
minima at $\pm \vev$ is lifted by the supersymmetry breaking
cubic term on the brane (\ref{couplings_susy}). The induced energy 
splitting between the vacua is then of order 
$\Delta V\sim E\vev^3\sim k^4\Omega^{-10-3a^2}$.
The preferred value of the cosmological constant (\ref{lambda})
then leads to $a^2\sim -2/3$, i.e. the scalar field is
localized somewhat towards the TeV-brane. The mass
of the scalar particles associated with $\Phi$ is of
order $\vev\sim k\Omega^{-5/3}\sim 10^2$ eV.

\section{Quantum corrections}
The case with a vanishing bulk mass discussed in 
section 3, can readily be embedded in the supersymmetric 
framework by setting $a=0$ in eq.~(\ref{couplings_susy}).
Then supersymmetry helps to tame radiative corrections. 
As in four dimensions the soft mass of the scalar zero mode, $m^2$, 
will receive a 1-loop quantum correction  
$\delta m^2\sim(1/16\pi^2)m^2\lambda\ln(\Lambda)$
by the exchange of the
scalar zero mode and its superpartner. Here $\Lambda$ 
denotes the momentum cut-off and $\lambda$ stems from the 
self coupling localized at Planck-brane. As discussed above, in the case
$a=0$ we have $m^2\sim k^2\Omega^{-4}$ and $\lambda\sim 1$.
In the warped model there are additional radiative corrections
to $m^2$ by the exchange of KK states of the scalar field.
The KK states are localized towards the TeV-brane. Therefore
they acquire a larger soft mass of order $k^2\Omega^{-2}$.
The quartic coupling between two zero modes and two KK states
of the scalar field is of order $\Omega^{-2}$. Thus loops
with  KK states of the scalar are of the same order as
zero mode loop. It remains to be seen if performing the
sum over the KK contributions, which are individually small,
leads to a destabilization of the soft mass of the zero mode.  

Since we assume the scalar field to be a gauge singlet, further
quantum corrections can only come from gravity. Exchange of
a zero mode gravitino in the loop gives rise to 
$\delta m^2\sim(1/16\pi^2)m^2_{3/2}(\Lambda/M_{\rm Pl})^2$,
where $m_{3/2}$ is the gravitino mass. Since for the zero mode
of the gravitino $m_{3/2}\sim k\Omega^{-3}$, we obtain a
tiny correction even for a Planck-size cut-off. For the KK
gravitinos the situation is different. Being localized towards the
TeV-brane, their supersymmetry breaking mass is TeV-size $(k\Omega^{-1})$
\cite{GP2} and their dimension-six interaction with the scalar
zero mode is suppressed by $1/(\Omega M_{\rm Pl}^2)$. The
corresponding correction to the scalar mass is then 
$\delta m^2\sim(1/16\pi^2)k^2\Omega^{-3}(\Lambda/M_{\rm Pl})^2$.
Since the dimension-six operator has its support at the TeV-brane,
it seems plausible that the KK gravitino loop is cut off at the TeV-scale.
In this case its contribution to the scalar mass 
$\delta m^2\sim(1/16\pi^2)k^2\Omega^{-5}$ is safely suppressed. 
However, given the limitations and uncertainties which are inherent 
in our treatment of radiative corrections, the issue of the quantum
stability of our approach to the dark energy problem is not yet
satisfactorily settled.

\section{Lifetime of the False Vacuum} 
The false vacuum state is metastable and will finally relax 
to the true vacuum by thermal fluctuations and/or quantum tunneling.
We must require the lifetime of the false 
vacuum to be longer than the age of the universe.
Because of the low temperature of the present universe, quantum 
tunneling is the dominant decay process. In the WKB approximation the
tunneling probability is given by \cite{tunnel}
\begin{equation} \label{prob}
p\sim\left(\frac{t_U}{R}\right)^4e^{-B},
\end{equation}
where $t_U$ denotes the age of the universe, and $R\sim1/ m$ 
is the characteristic scale of the problem. In the thin wall
approximation the Euclidean action of the tunneling configuration
\begin{equation}
B=27\pi^2\frac{S_1^4}{2(\Delta V)^3},
\end{equation}
depends on the energy splitting of the vacua $\Delta V$ and
the ``surface tension'' of the bubble $S_1=\int d\phi\sqrt{2V(\phi)}$. 

Vacuum decay is most effective between minima which are
close to each other in field space and separated only by a
low energy barrier $V_B$. Among the examples discussed before
the tunneling rate is largest in the case  of section 3 where
$m\sim\vev\sim 10^{-3}$ eV. From the constraint $p<1$
we obtain $B\gsim 280$.  For a polynomial potential one
finds $B\approx 7\cdot10^3 (V_B/\Delta V)^3/\lambda$, 
i.e.~the false vacuum is metastable even for $V_B\sim \Delta V$,
as we expect for a potential with only a single mass scale.
In the other example we discussed, the barrier heights and
scalar vevs are even much larger. Thus, in all the cases discussed
the false vacuum is sufficiently long lived.

\section{Conclusions}
In conclusion, we have shown that a warped geometry setting could help provide
a new explanation of the origin of the tiny vacuum energy density of the
universe that is indicated by recent observations. This can either be in the
form of a cosmological constant or, if supersymmetry is invoked, as contribution
arising from a slow rolling quintessence field. Although no explanation exists
as to why all other contributions to the vacuum energy density effectively vanish,
it is certainly intriguing to think that resolution of the gauge hierarchy 
and dark energy problems may have a common origin. 

\section*{Acknowledgements}
We acknowledge useful discussions with A.~Hebecker and A.~Pomarol
 and a helpful correspondence with T. Gherghetta.
We wish to thank the Alexander von Humboldt Stiftung for providing the
impetus for our collaboration. Q.S.~also acknowledges the hospitality
of the Theory Groups at DESY and Heidelberg, especially Wilfried
Buchm\"uller, Michael Schmidt and Christof Wetterich. This work was 
supported in part by DOE under contract DE-FG02-91ER40626.


\begin{thebibliography}{12}
\bibitem{CC_OBS}S.~Perlmutter {\em et al.}, {\em Ap.~J.} {\bf 483} (1997) 565;
                S.~Perlmutter {\em et al.}, astro-ph/9812473;             
                B.~Schmidt {\em et al.}, {\em Ap.~J.} {\bf 507} (1998) 46;
                A.J.~Riess {\em et al.}, {\em A.~J.} {\bf 116} (1998) 1009;
                astro-ph/0104455.

\bibitem{EV} For recent reviews of various suggestions for resolving
                   the cosmological constant problems and additional references,
                   see  A.~Vilenkin, hep-th/0106083;   
                  U.~Ellwanger,  hep-ph/0203252. 
\bibitem{KKS}S.~Kachru, J.~Kumar and E.~Silverstein, {\em Phys. Rev.} {\bf D59}
             (1999) 106004; 
             E.I.~Guendelman, {\em Mod. Phys. Lett.} {\bf A14} (1999) 1043;
             P.H.~Frampton, hep-th/0002053; 
             N.~Arkani-Hamed, L.J.~Hall, C.~Colda and Murayama,  {\em Phys. Rev. Lett.} 
             {\bf 85} (2000) 4434;
             S.M.~Barr and D.~Seckel, {\em Phys. Rev.} {\bf D64} (2001) 123513;
             T.~Banks and M.~Dine, {\em JHEP} {\bf 0110} (2001) 012.
\bibitem{RS} L.~Randall and R.~Sundrum, {\em Phys. Rev. Lett.} 
             {\bf 83} (1999) 3370.

\bibitem{G} M.~Gogberashvili, hep-ph/9812296. 

\bibitem{GN} Y.~Grossman and M.~Neubert, 
                     {\em Phys. Lett.} {\bf B474} (2000) 361 [hep-ph/9912408].

\bibitem{GP} T.~Gherghetta and A.~Pomarol, 
               {\em Nucl. Phys.} {\bf B586} (2000) 141  [hep-ph/0003129].

\bibitem{HS2} S.J.~Huber and Q.~Shafi, {\em Phys. Lett.} {\bf B498} (2001) 256 
                    [hep-ph/0010196]. 

\bibitem{HS3}S.J.~Huber and Q.~Shafi, {\em Phys. Lett.} {\bf B521} (2001) 365 
                    [hep-ph/0104293]; hep-ph/0205327. 

\bibitem{TWCF}S.H.~Henry Tye and I.~Wasserman {\em Phys. Rev. Lett.} {\bf 86}
                      (2001) 1682 [hep-th/0006068];
                      J.M.~Cline and H.~Firouzjahi, {\em Phys. Lett.} {\bf B514} (2001) 205
                      [hep-ph/0012090]. 

\bibitem{quin}C.~Wetterich, {\em Nucl. Phys.} {\bf B302} (1988) 668;
                    P.J.E.~Peebles and B.~Ratra, {\em Ap. J. Lett.} {\bf 325} (1988) L17;
                    R.R~Caldwell, R.~Dave and P.J~Steinhard, {\em Phys. Rev. Lett} 
                    {\bf 80} (1998) 1582; 
                    I.~Zlatev, L.~Wang and P.J~Steinhard, {\em Phys. Rev. Lett} 
                    {\bf 85} (2000) 4434. 
                
\bibitem{GW} W.D.~Goldberger and M.B.~Wise, 
                     {\em Phys.~Rev.} {\bf D60} (1999) 107505 [hep-ph/9907218]; 
                     {\em Phys. Rev. Lett.} {\bf 83} (1999) 4922 [hep-ph/9907447]. 

\bibitem{HS}S.J.~Huber and Q.~Shafi, {\em Phys. Rev.} {\bf D63} (2001) 045010  
                    [hep-ph/0005286].

\bibitem{GP2}T.~Gherghetta and A.~Pomarol, {\em Nucl.~Phys.} {\bf B602} (2001) 3                                   [hep-ph/0012378]. 

\bibitem{DV}G.R.~Dvali and A.~Vilenkin, {\em Phys. Rev.} {\bf D64} (2001) 063509 
                 [hep-th/0102142]. 

\bibitem{tunnel}S.~Coleman,  {\em Phys.~Rev.} {\bf D15} (1977) 2929;
                       S.~Coleman and C.~Callan,  {\em Phys.~Rev.} {\bf D16} (1977) 1762.


\end{thebibliography}
\end{document}